\documentclass[preprint,12pt]{elsarticle}

\usepackage{amssymb}
\usepackage{epsfig,bm,feynmf}
\usepackage{graphics}

\journal{Physica B}

\begin{document}

\begin{frontmatter}



\title{Intrinsic Spin Hall Effect with Spin-Tensor-Momentum Coupling}

\author{Yong-Ping Fu$^{1,*}$, Fei-Jie Huang$^2$ and Qi-Hui Chen$^{3,**}$}

\address{$^1$Department of Physics, West Yunnan University, Lincang 677000, China}
\address{$^2$Department of Physics, Kunming University, Kunming 650214, China}
\address{$^3$School of Physical Science and Technology, Southwest Jiaotong University, Chengdu 610031, China}
\address{Corresponding author: $^*$Email: ynufyp@sina.cn; $^{**}$Email: qhchen@swjtu.edu.cn}

\begin{abstract}
We derive the spin continuity equation by using the Noether's theorem. A new type of spin-tensor Hall current is found in the continuity equation. The spin-tensor Hall current is originating from the coupling of the spin-tensor and the momentum. The intrinsic spin Hall effect in the two-dimensional fermion model with the spin-orbit coupling and spin-tensor-momentum coupling is studied. The total spin Hall conductivity with the presence of the spin-tensor Hall current is calculated. The numerical results indicate that the total spin Hall conductivity is enhanced by the contribution of the spin-tensor-momentum coupling. The spin-tensor-momentum coupling may increase the spin transport in the intrinsic spin Hall effect.
\end{abstract}




\end{frontmatter}



\section{Introduction}

The spin-orbit coupling (SOC) plays an important role in many quantum phenomena of solid and ultracold atomic systems. One important example is the spin Hall effect (SHE) \cite{rev1}. The SHE is studied in the semiconductor heterostructures. The spin Hall current is independent of the skew scattering of the moving magnetic moments and the magnetic impurities \cite{S C Zhang1}. In the SHE a dissipationless spin Hall current can be induced by an electric field in hole-doped semiconductors \cite{S C Zhang2,S C Zhang3,hole1}. The SHE has been observed experimentally in GaAs \cite{exp1,exp2,exp3}. The effect does not depends on the electron-impurity scattering \cite{ESHE0,ESHE1,ESHE2,ESHE3,ESHE4}, but is related to the substantial SOC. In the momentum ($p$) direction perpendicular to the electric current, the effective torque from the Rashba SOC tilts the spins up for $p>0$ and down for $p<0$. The effect is intrinsic in electron systems, and is called as intrinsic spin Hall effect (ISHE). The spin Hall conductance is found to be a universal value, and is independent of the Rashba coupling strength \cite{SHE2,SHE4,SHE1,SHE5,SHE3}.

Recently, a new type of spin-tensor-momentum coupling (STMC) is proposed in the Bose-Einstein condensation \cite{STMC1}. It is found that the coupling between the two components of particle's spin and momentum (e.g., orbit) leads to new types of stripe superfluid phase and multicritical points for phase transitions. The STMC may open a door for exploring many other interesting physics \cite{STMC11,Mas0,Mas1,Mas2,Mas3,Mas4}. For example, the new properties of the SHE in the STMC system. If a physical system depends on two spin degrees of freedom, the concept of spin-tensor can be used to study this physical problem. The electric current depends only on the velocity degree of freedom, while the spin current depends on the velocity and spin degrees of freedom. The spin-tensor current depend on velocity and spin-tensor degrees of freedom, the STMC will inevitably lead to more abundant physical phenomena with the increasing of the degree of freedom.

This paper is organized as follows. We discuss the spin continuity equation by using the Noether's theorem, and define the new spin-tensor Hall current in Sec.2. In Sec.3 we study the ISHE in a model with the Rashba SOC and STMC. The conductivities of the spin Hall and spin-tensor Hall currents are calculated. Finally the conclusion is presented in Sec.4.

\section{Spin Continuity Equation and Spin-Tensor Hall Current}
Noether's theorem indicates that the invariant of the system under a continuous transformation will lead to the corresponding conservation current \cite{nother1}. One can rigorously derive the total-angular momentum conservation equation by using the Noether's theorem. The spin current of the Dirac fermion coupled with the external electromagnetic field can be extracted from total-angular momentum current (Appendix A). The non-relativistic approximation of the spin current can be obtained by using the standard Foldy-Wouthuysen (FW) transformation \cite{FW1,RQM}. The spin continuity equation up to the order of $1/c^{2}$ reads
\begin{eqnarray}
\frac{\partial}{\partial t}\rho_{s}^{i}
+\nabla_{j}J_{s}^{ji}=T^{i}_{s} ,
\label{conser01}
\end{eqnarray}
the spin density is
\begin{eqnarray}
\rho_{s}^{i} \!= \!\psi^{+}s
^{i}\psi\! +\!\psi ^{+}\!\frac{1 }{4c^{2}}\!\left\{\!\left[\boldsymbol{v}\!\!\times\!\!(\boldsymbol{v}\!\!\times\!\!
\boldsymbol{s})\right]^{i}
\!-\!\left[(\boldsymbol{v}\!\!\times\!\!
\boldsymbol{s})\!\!\times\!\!\boldsymbol{v}\right]^{i}\!\right\}\psi,
\label{spindensity01}
\end{eqnarray}
where $\boldsymbol{s}=\hbar \boldsymbol{\sigma}/2$ is the spin operator, $\boldsymbol{v}=\boldsymbol{\pi} /m$ is the velocity operator, $\boldsymbol{\pi}=\boldsymbol{p}-e\boldsymbol{A}/c$ is the momentum operator. The second term of the spin density is the relativistic correction up to $1/c^{2}$ order and is very small. The correction is the contribution from the carrier's spin coupling to its momentum. It includes the spin-vector potential interaction $\boldsymbol{A}\times\boldsymbol{s}$ and the spin-orbit coupling $\boldsymbol{p}\times\boldsymbol{s}$.

The spin current can be derived as
\begin{eqnarray}
&&J_{s}^{ji} \!= \!\psi^{+}\!\frac{1}{4}\!\left(\!s^{i}v^{j}\!-\!s^{j}v^{i}
\!\right)\!\psi\!
+ \!\psi^{+}\!\frac{i}{2}\!\left[\!s^{i}\!(\!\boldsymbol{\tau}\!\!\times\!\!\boldsymbol{v}\!)^{j}
\!-\!s^{j}(\boldsymbol{\tau}\!\!\times\!\!\boldsymbol{v})^{i}\!
\right]\!\psi \nonumber\\
&&+\psi^{+}\!\frac{1}{2}\!\delta^{ij}\!\boldsymbol{s}\!\cdot\!\boldsymbol{v}\!\psi
-\psi^{+}\!\frac{1}{12c^{2}}\!\left( J^{ji}_{s(0)}\boldsymbol{v}^{2}+ \boldsymbol{\sigma}\cdot\boldsymbol{v}J^{ji}_{s(0)}\boldsymbol{\sigma}\cdot\boldsymbol{v}\right)\!\psi \nonumber\\
&&+\mathrm{H.c.},
\label{spincurrent01}
\end{eqnarray}
here we define $\boldsymbol{\tau}= \boldsymbol{\sigma}/2$ as the pseudospin and  $J^{ji}_{s(0)}\!\!=\!\!\left(\!s^{i}v^{j}\!-\!s^{j}v^{i}
\!\right)\!/\!4
\!+\!i \left[\!s^{i}\!(\!\boldsymbol{\tau}\!\!\times\!\!\boldsymbol{v}\!)^{j}
\!-\!s^{j}(\boldsymbol{\tau}\!\!\times\!\!\boldsymbol{v})^{i}\!
\right]/2$. The first term of Eq.(\ref{spincurrent01}) is the spin Hall current
\begin{eqnarray}
J^{ji}_{\mathrm{SH}}=\!\psi^{+}\!\frac{1}{4}\!\left(\!s^{i}v^{j}\!-\!s^{j}v^{i}
\!\right)\!\psi+\!\mathrm{H.c.}.
\label{spincurrent011}
\end{eqnarray}
Under the time reversal transformation $t \rightarrow-t$, we have $\boldsymbol{s}\rightarrow-\boldsymbol{s}$, $\boldsymbol{\pi}\rightarrow-\boldsymbol{\pi}$, the spin Hall current satisfies the time-reversal-symmetry. The response equation of the spin Hall current $J^{ij}_{\mathrm{SH}}=\sigma_{\mathrm{SH}}\varepsilon^{ijk}E_{k}$ \cite{S C Zhang1,S C Zhang2} requires the antisymmetry of $i$ and $j$ components, here $\sigma_{\mathrm{SH}}$ is the spin Hall conductivity. The same antisymmetry structure appears in the Eq.(\ref{spincurrent011}).

We define the second term of Eq. (\ref{spincurrent01}) as the spin-tensor Hall current
\begin{eqnarray}
J_{\mathrm{STH}}^{ji} \!=\!\psi^{+}\!\frac{i}{2}\!\left[\!s^{i}\!(\!\boldsymbol{\tau}\!\!\times\!\!\boldsymbol{v}\!)^{j}
\!-\!s^{j}(\boldsymbol{\tau}\!\!\times\!\!\boldsymbol{v})^{i}\!
\right]\!\psi+\!\mathrm{H.c.}.
\label{spincurrent02}
\end{eqnarray}
Mathematically, the rank-2 spin-tensor is defined as $N_{ij}=P_{ij}+D_{ij}$, where $P_{ij}=(S_{i}S_{j}+S_{j}S_{i})/2$ is the symmetry tensor and $D_{ij}=(S_{i}S_{j}-S_{j}S_{i})/2$ is the antisymmetry tensor. $S_{i}$ denotes the fermion spin $s_{i}$ or pseudospin $\tau_{i}$. In Eq. (\ref{spincurrent02}) $s^{i}(\boldsymbol{\tau}\times\boldsymbol{v})^{j}=s^{i}\tau^{k} v^{l}\varepsilon^{klj}$ denotes the coupling of the spin-tensor and the momentum, where $s^{i}\tau^{k}$ ($i\neq j\neq k $) is the antisymmetry spin-tensor. The spin-tensor introduced in Ref. \cite{STMC1} applies to systems with spin $s$=1. The spin operator of the boson is not satisfied with the commutation relation. Therefore the spin-tensor of Ref. \cite{STMC1} is a kind of symmetry tensor.

The spin-tensor Hall current is originating from the coupling of the spin-tensor and the momentum. We note that the spin-tensor Hall current is not a higher order correction of the non-relativistic approximation, it is in the same order as the spin Hall current. The imaginary number $i$ in Eq.(\ref{spincurrent02}) is to ensure that the spin-tensor current operator [$is^{i}(\boldsymbol{\tau}\times\boldsymbol{v})^{j}$] is the Hermite current operator. In the time-reversal transformation $\boldsymbol{s}\rightarrow-\boldsymbol{s}$ and $\boldsymbol{\pi}\rightarrow-\boldsymbol{\pi}$ the spin-tensor current is also time-reversal symmetry protected. Because of the pseudospin $\boldsymbol{\tau}$ does not contain the constant $\hbar$, where the dimension of $\hbar$ is [J$\cdot$ s], the pseudospin is dimensionless, and therefore in the time-reversal transformation we have $\boldsymbol{\tau}\rightarrow \boldsymbol{\tau}$. It is also easy to proved that the spin-tensor current is antisymmetric current for $i$ and $j$ component. These properties indicate that the second term of Eq.(\ref{spincurrent01}) is a kind of spin-tensor Hall current.

A new velocity operators can be obtained by analogy with the conventional spin Hall current. We define the velocity of the spin-tensor current as $\boldsymbol{v}_{\mathrm{ST}}=2(\boldsymbol{\tau}\times\boldsymbol{v})$, and therefore the total spin Hall current can be defined as
\begin{eqnarray}
J_{\mathrm{tot-SH}}^{ji} \!= \!\psi^{+}\!\frac{1}{4}\!\left(\!s^{i}v_{\mathrm{tot}}^{j}\!-\!s^{j}v_{\mathrm{tot}}^{i}
\!\right)\!\psi+\!\mathrm{H.c.},
\label{spincurrent04}
\end{eqnarray}
where the total velocity is
$\boldsymbol{v}_{\mathrm{tot}}\!=\!\boldsymbol{v}\!+\!i\boldsymbol{v}_{\mathrm{ST}}$.

The third term of Eq.(\ref{spincurrent01}) is the scalar current, and $\delta^{ij}\psi^{+}\left(\boldsymbol{s}\cdot\boldsymbol{v}\right)\psi=0$ when $i\neq j$.
The fourth term of Eq.(\ref{spincurrent01}) includes the operators of spin Hall current and spin-tensor Hall current. This term is a kind of relativistic correction from the SOC and STMC. The relativistic correction is of the $1/c^{2}$ order and is small enough to be neglected.

In the external electromagnetic field, the total spin current is not conserved because of the coupling of the angular momentum and external field. The non-relativistic limit of the spin torque (up to $1/c^{2}$ order) can be derived in the form
\begin{eqnarray}
&&T^{i}_{s} \!=\!\psi^{+}\!\!\frac{e}{mc}\!\left(\boldsymbol{s}\!\times\!\boldsymbol{B}\right)\!\psi -\!\psi^{+}\!\frac{e\hbar^{2}}{4m^{2}c^{2}}\left(\nabla\times \boldsymbol{E}\right)\!\psi \nonumber\\
&&+\!\psi^{+}\!\!\frac{e}{2m^{2}c^{2}}\left\{\!\left[\boldsymbol{\pi}\!\times\!(\boldsymbol{E}\! \times \! \boldsymbol{s})\right]^{i}\!+\!\left[(\boldsymbol{s}\! \times \! \boldsymbol{E})\!\times\!\boldsymbol{\pi}\right]^{i}\!\right\}\!\psi,  
\label{torque01}
\end{eqnarray}
where $\boldsymbol{B}$ is the magnetic field and $\boldsymbol{E}$ is the electric field. We find that the spin torque is the result of the relativistic corrections. The first term of Eq.(\ref{torque01}) is the coupling of the magnetic moment $\boldsymbol{\mu}(=e\boldsymbol{s}/mc)$ and the external field $\boldsymbol{B}$. The second term of the torque is the contribution from curl$ \boldsymbol{E}$. The second term equals zero when the external electric field $\boldsymbol{E}$ is constant in the SHE. The third term in $T^{i}_{s}$ is the contribution of $\boldsymbol{\pi}\times\boldsymbol{v}'$ and is finite for the electric orbital motion in the presence of SOC \cite{NR3}, where $\boldsymbol{v}'=(e/2m^{2}c^{2})(\boldsymbol{s} \times \boldsymbol{E})$ is the relativistic correction of the velocity. If the external field $\boldsymbol{B}=0$, the spin torque is of the order $1/c^{2}$, the spin current is approximately conserved.

Because of the coupling of the spin and orbit, the contribution of spin angular momentum and the orbit angular momentum to the torque can not be directly divided. Both the spin and orbit angular momentums contribute to the spin torque. In Ref. \cite{NR3} the authors ignore the total contribution of the orbit angular momentum to the spin torque. The authors in Ref. \cite{NR1} use the Gordon decomposition to divide the spin current into the convective and internal parts, and derive the continuity equation for the convective spin current with the absence of the external electromagnetic field. They consider influence of the external electromagnetic field by using the simple transformation $\boldsymbol{p}\rightarrow \boldsymbol{\pi}$ and $i\hbar\partial/\partial t\rightarrow i\hbar\partial/\partial t-eA_{0}$ in the final results. In our derivation the external field $A_{\mu}$ is included in the total-angular momentum conservation equation and the FW transformation. Thus the results of the continuity equation of the spin current is more valid in this paper.

\section{ Spin Hall and Spin-Tensor Hall Conductivities in ISHE}
\subsection{Model Hamiltonian }

In the two-dimensional fermion system (2DFS) the substantial Rashba SOC and the external electric field $E$ lead to the spin Hall current \cite{SHE2,SHE4,SHE1,SHE3,NR2}. The Hamiltonian including the SOC and STMC is given by
\begin{eqnarray}
H_{0}=\frac{\boldsymbol{p}^{2}}{2m}+H_{\mathrm{R}}+H_{\mathrm{ST}},
\label{Hamilton1}
\end{eqnarray}
where $H_{\mathrm{R}}=-\frac{2\lambda}{\hbar}(s_{y}p_{x}-s_{x}p_{y})$ is the Rashba SOC Hamiltonian, $\lambda$ is the Rashba coupling constant \cite{Rashba}. In the $x-y$ plane the Rashba coupling of the spin $\boldsymbol{s}$ and the momentum $(p_{x},p_{y})$ is $(\boldsymbol{s}\times\boldsymbol{p})_{z}=s_{x}p_{y}-s_{y}p_{x}$, where the Rashba SOC term $s_{x}p_{y}$ describes the coupling between the linear momentum $p_{y}$ and the spin $s_{x}$ in the $p_{y}$ momentum direction. In the $p_{x}$ momentum direction, the Rashba SOC can be described as $s_{y}p_{x}$. We define the Rashba-like STMC as $\mathcal{T}_{zx}p_{y}$ and $\mathcal{T}_{zy}p_{x}$ in the $(p_{x},p_{y})$ momentum space, where $\mathcal{T}_{ij}=\tau_{i}s_{j}$ is the spin-tensor.

The spin-tensor Hall current will not appear in the 2DFS with the absence of the STMC. In order to induce the spin-tensor Hall current in 2DFS, we construct a Hamiltonian including the SOC and STMC
\begin{eqnarray}
H_{\mathrm{ST}}=\frac{4\zeta}{m\hbar^{2}} \left[(\mathcal{T}_{zy}p_{x})(s_{y}p_{x})-(\mathcal{T}_{zx}p_{y})(s_{x}p_{y} )\right],
\label{Hamilton2}
\end{eqnarray}
and we define $\zeta$ as the dimensionless coupling constant of the STMC, and $0<\zeta< 1$. The physical origin of the pseudospin $\tau_{i}$ can be the double sublattices structure of the 2DFS with SOC. After simple algebraic operations, the energy states of Eq. (\ref{Hamilton2}) can be written as $E_{\mathrm{ST}}=\pm\zeta \cos 2\theta p^{2}/2m$, where $\theta=arc \tan (p_{y}/p_{x})$. The pseudospin $\langle\tau_{z}\rangle=\pm 1/2$ corresponds to the two sublattice energy states.

\begin{figure}[ptb]
\includegraphics[width=0.7\linewidth]{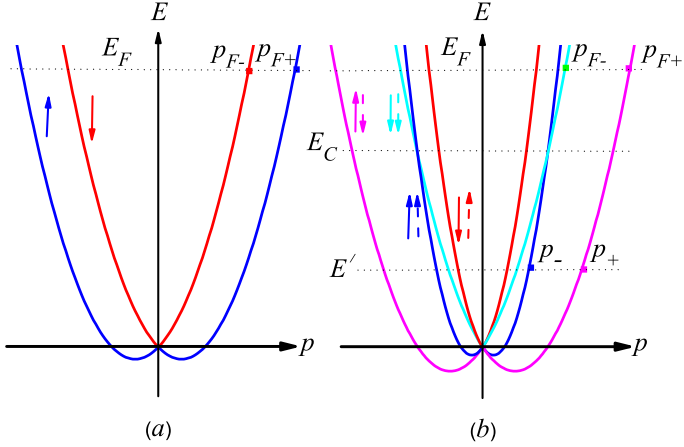}
\caption{Schematic of the splitting of energy states due to the Rashba SOC (a) and SOC+STMC (b). Solid and dash arrow stand for spin and pseudospin, respectively. SOC+STMC leads to four eigenstates of the 2DFS. We take the eigenvalues as $\langle s \rangle=\hbar/2$(up arrow), $-\hbar/2$(down arrow), and $\langle\tau_{z}\rangle=1/2$(up arrow), $-1/2$(down arrow). \label{fig1}}
\end{figure}

The Rashba SOC splits energy states into two branches (Fig.\ref{fig1}(a))
\begin{eqnarray}
E_{\eta}=\frac{p^{2}}{2m}-\eta \lambda p,
\end{eqnarray}
where $\eta=+1$ for $\langle s\rangle=\hbar/2$ and $\eta=-1$ for $\langle s\rangle=-\hbar/2$. The SOC and STMC lead to four eigenstates (Fig.\ref{fig1}(b))
\begin{eqnarray}
E_{\eta,\eta'}=\frac{p^{2}}{2m}\left(1+\eta'\zeta \cos 2\theta\right)- \eta\lambda p,
\end{eqnarray}
where $\eta'=+1$ for $\langle \tau_{z}\rangle=1/2$, $\eta'=-1$ for $\langle \tau_{z}\rangle=-1/2$, and $\tan \theta=p_{y}/p_{x}$ with the momentums $p_{x}=p\cos \theta$ and $p_{y}=p\sin \theta$. The SOC+STMC Hamiltonian has four eigenstates, the fermions occupy the lower energy band $E_{+,-}$ in the spin Hall effect. The fermi energy $E_{F}$ is much larger than the slitting energy $\Delta$ \cite{split1}, we assume $E_{F}>E_{c}\gtrsim\Delta$, where $E_{c}$ is the cross point of the energy bands $E_{+,+}$ and $E_{-,-}$. The difference of the Fermi radii $p_{F+}$ and $p_{F-}$ of the majority ($E_{+,-}$) and minority ($E_{-,-}$) spin bands can be calculated by the eigenvalues
\begin{eqnarray}
E_{F}\!\!&&=\frac{p^{2}_{F+}}{2m}(1-\zeta \cos 2\theta)- \lambda p_{F+} \nonumber\\
&&=\frac{p^{2}_{F-}}{2m}(1-\zeta \cos 2\theta)+ \lambda p_{F-},
\end{eqnarray}
and we have
\begin{eqnarray}
\triangle p_{F}=p_{F+}-p_{F-}=\frac{2m\lambda}{1-\zeta \cos 2\theta}.
\end{eqnarray}
If the Fermi energy is smaller than the energy $E_{c}$, such as $E'_{F}=E'$, the Fermi space is between the energy bands $E'_{+,+}(p'_{F-})$ and $E'_{+,-}(p'_{F+})$. The difference $\triangle p'_{F}$ of the Fermi radii $p'_{F+}$ and $p'_{F-}$ depends on the Fermi energy $E'_{F}$.

\subsection{Spin Hall and Spin-Tensor Hall Conductivities }
In this section we discuss the spin Hall current which is polarized in the $\hat{z}$-direction and flows in the $\hat{y}$-direction, and is perpendicular to the charge current $\hat{x}$-direction. Based on Eq.(\ref{spincurrent04}) the total spin Hall current operator is defined as $\hat{J}^{yz}_{\mathrm{tot-SH}}=\frac{1}{4}(s_{z}v_{\mathrm{tot}y}-s_{y}v_{\mathrm{tot}z})+\mathrm{H.c.}=
\frac{1}{4}(\{s_{z},v_{\mathrm{tot}y}\}-\{s_{y},v_{\mathrm{tot}z}\})$, here $v_{\mathrm{tot}z}=0$. We define the velocities according to the Heisenberg equations
\begin{eqnarray}
v_{y}=\frac{1}{i\hbar}[y,\frac{\boldsymbol{p}^{2}}{2m}+H_{\mathrm{R}}],
\label{v1}
\end{eqnarray}
and
\begin{eqnarray}
v_{\mathrm{ST}y}=\frac{1}{i\hbar}[y,H_{\mathrm{ST}}].
\label{v2}
\end{eqnarray}
The velocity operators of the spin current and the spin-tensor current can be calculated as $v_{y}\!=\!\frac{p_{y}}{m}\!+\!\frac{2\lambda}{\hbar} s_{x}$ and $v_{\mathrm{ST} y}\!=\!-2\zeta\tau_{z}\frac{p_{y}}{m}$. In the SOC+STMC 2DFS the pseudospin and spin have the same index $z$,  we redefine the total velocity operator as $v_{\mathrm{tot}y}=v_{y}+v_{\mathrm{ST}y}$ to ensure the Hermiticity of the spin current.

The stable spin Hall effect requires that the spin is along a certain direction. However the spin is precessing with time due to the SOC, the precession equation of the spin can be obtained by the Heisenberg equations ($ds_{i}(t)/dt=[s_{i},H]/i\hbar$) as 
\begin{eqnarray}
\frac{ds_{1}(t)}{dt}=-\frac{2\lambda}{\hbar} s_{z}(t)p_{1},
\label{spinmotion1}
\end{eqnarray}
\begin{eqnarray}
\frac{ds_{2}(t)}{dt}=-\frac{2\lambda}{\hbar} s_{z}(t)p_{2},
\label{spinmotion2}
\end{eqnarray}
\begin{eqnarray}
\frac{ds_{z}(t)}{dt}=\frac{2\lambda}{\hbar}  \left[ s_{1}(t)p_{1}+s_{2}(t)p_{2} \right],
\label{spinmotion3}
\end{eqnarray}
where $\hat{x}_{1}$ and $\hat{x}_{2}$ denote the azimuthal and radial direction in momentum space, respectively.

We consider the presence of the external electric field, the system Hamiltonian is given by $H=H_{0}+eE_{x}x$, where $E_{x}$ is the external electric field in the $\hat{x}$-direction. Using the Heisenberg equation, we have $p_{x}=p_{x0}-eE_{x}t$ where $p_{x0}$ is the initial momentum at $t=0$. The fermi surface is displaced an amount $eE_{x}t$ due to the presence of the external electric field. Applying the adiabatic spin dynamics \cite{SHE2}, the $\hat{x}_{2}$ component of the spin can be approximated to $s_{2}(t)=s\sin \theta p_{1}(t)/p_{2}$ for a weak field $E_{x}$ and a short instant $t$. Substituting the above approximation in Eq.(\ref{spinmotion2}) we have the $\hat{z}$ component of the spin $s$ as
\begin{eqnarray}
s_{z}=-\frac{\hbar}{2\lambda p_{2}}\frac{ds_{2}(t)}{dt}.
\label{spinmotion2-2}
\end{eqnarray}
In the weak field and short instant approximation, $\dot{p}_{1}(t)\approx \dot{p}_{x}(t)=-eE_{x}$ and $p_{2}\approx p$, the electric field induces a linear response of spin $s_{z}$:
\begin{eqnarray}
s_{z}=\frac{e\hbar s \sin\theta}{2\lambda p^{2}}E_{x}.
\label{spinmotion2-3}
\end{eqnarray}
The SOC+STMC Hamiltonian has four eigenstate, just the lower energy band $E_{+,-}$(majority spin band) contributes to the total spin current, here the values of the spin and the pseudospin are $\langle s\rangle=\hbar/2$ and $\langle \tau_{z}\rangle=-1/2$.

From the definition of spin Hall and spin-tensor Hall current, the corresponding current operators are given by $\hat{J}^{yz}_{\mathrm{SH}}=\frac{1}{4}\left(\{s_{z},v_{y}\}-\{s_{y},v_{z}\}\right)$ and $\hat{J}^{yz}_{\mathrm{STH}}=\frac{1}{4}\left(\{s_{z},v_{\mathrm{ST}y}\}-\{s_{y},v_{\mathrm{ST}z}\}\right)$, where $v_{ z}=v_{\mathrm{ST} z}=0$ in this 2DFS. The definition of the spin Hall current is different from the conventional definition of the spin current ($\hat{J}^{yz}_{\mathrm{spin}}=\frac{1}{2}\{s_{z},v_{y}\}$), because the conventional spin current operator can be divided into two parts $\hat{J}^{yz}_{\mathrm{spin}}=\hat{J}^{yz}_{\mathrm{spin(1)}}+\hat{J}^{yz}_{\mathrm{spin(2)}}$, where
\begin{eqnarray}
\hat{J}^{yz}_{\mathrm{spin(1)}}=\frac{1}{4}\left(\{s_{z},v_{y}\}-\{s_{y},v_{z}\}\right),
\label{con-spincurrent1}
\end{eqnarray}
and
\begin{eqnarray}
\hat{J}^{yz}_{\mathrm{spin(2)}}=\frac{1}{4}\left(\{s_{z},v_{y}\}+\{s_{y},v_{z}\}\right).
\label{con-spincurrent2}
\end{eqnarray}
The antisymmetry part $\hat{J}^{yz}_{\mathrm{spin(1)}}$ is the spin Hall current operator. The operator $\hat{J}^{yz}_{\mathrm{spin(2)}}$ does not satisfy the antisymmetry structure of indexes $y$ and $z$, and therefore $\hat{J}^{yz}_{\mathrm{spin(2)}}$ is a kind of non-spin-Hall current operator. Only the antisymmetry part of the spin current contributes to the SHE \cite{S C Zhang2,NR3}. It is inappropriate to define $\hat{J}^{yz}_{\mathrm{spin(1)}}$ as a spin Hall current operator in the SHE.

The spin Hall and spin-tensor Hall currents in the SOC+STMC system can be expressed as $J_{\mathrm{SH}}=\sum_{k}\langle \hat{J}_{\mathrm{SH}}\rangle f_{D}$ and $J_{\mathrm{STH}}=\sum_{k}\langle\hat{J}_{\mathrm{STH}}\rangle f_{D}$, where $k=p/\hbar$ is the wave vector and $f_{D}$ is the equilibrium Fermi-Dirac distribution. At zero temperature the spin current is given by
\begin{eqnarray}
J^{yz}_{\mathrm{SH}}&&=\int^{p_{F+}}_{p_{F-}}\!\!\!\frac{d^{2}p}{(2\pi\hbar)^{2}} \frac{\varepsilon_{xyz}e\hbar^{2} E_{x} \sin\theta p_{y}}{8\lambda m p^{2}}\nonumber\\
&&=\varepsilon_{xyz}\sigma_{\mathrm{SH}}E_{x},
\label{conductivity1}
\end{eqnarray}
where the spin Hall conductivity is
\begin{eqnarray}
\sigma_{\mathrm{SH}}=\frac{e}{16\pi^{2}}\int^{2\pi}_{0}d\theta \frac{\sin^{2}\theta}{1-\zeta\cos 2\theta}.
\label{conductivity2}
\end{eqnarray}
The spin-tensor Hall current is calculated by
\begin{eqnarray}
J^{yz}_{\mathrm{STH}}&&=\int^{p_{F+}}_{p_{F-}}\!\!\!\frac{d^{2}p}{(2\pi\hbar)^{2}} \frac{\varepsilon_{xyz}e\hbar^{2} \zeta E_{x}\sin\theta p_{y}}{8\lambda m p^{2}}\nonumber\\
&&=\varepsilon_{xyz}\sigma_{\mathrm{STH}}E_{x},
\label{conductivity3}
\end{eqnarray}
where the spin-tensor Hall conductivity is
\begin{eqnarray}
\sigma_{\mathrm{STH}}(\zeta)=\zeta\sigma_{\mathrm{SH}}(\zeta).
\label{conductivity4}
\end{eqnarray}
Eq.(\ref{conductivity2}) and (\ref{conductivity4}) indicates that the spin Hall and spin-tensor Hall conductivities are not universal values, they depend on the STMC parameter $\zeta$. If $\zeta=0$, the STMC becomes to zero ($\sigma_{\mathrm{STH}}=0$) and the spin Hall conductivity can be calculated as $\sigma_{\mathrm{SH}}=e/16 \pi$. The numerical results indicate that $\sigma_{\mathrm{STH}}\approx e/16\pi$ at $\zeta=1$. Sinova $et$ $al.$ found that the spin Hall conductivity in a system with Rashba SOC remains the universal value $e/8 \pi$ \cite{SHE2,SHE1,SHE3}. The two results of the spin Hall conductivities differ by only a factor $1/2$. This difference comes from the different definitions of the spin Hall current.

\begin{figure}[ptb]
\includegraphics[width=0.7\linewidth]{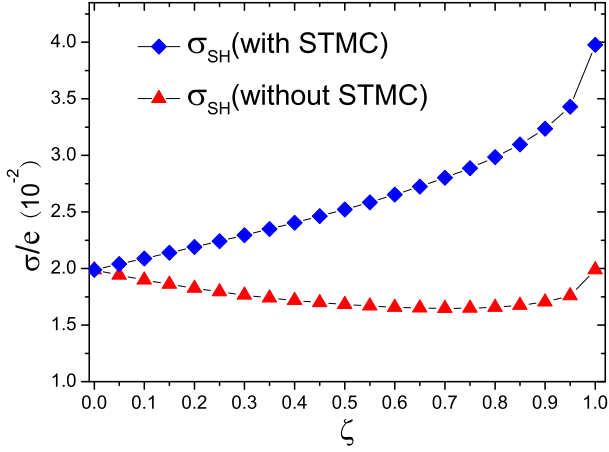}
\caption{The numerical results of the spin Hall conductivity vs the coupling strength $\zeta$. We have used the dimensionless quantity $\sigma/e$. The triangle symbol stands for the spin Hall conductivity of the conventional spin Hall current without the STMC. The diamond symbol stands for the total spin Hall conductivity of the total Hall current ($J_{\mathrm{SH}}+J_{\mathrm{STH}}$) with the STMC.
\label{fig2}}
\end{figure}

In Fig.\ref{fig2} we plot the spin Hall conductivities as a function of STMC strength in the condition of $E_{F}>E_{c}$. The spin Hall conductivity with the absence of the STMC is not a universal value, and we find $0.8 e/16\pi<\sigma_{\mathrm{SH}}<e/16\pi$ in the region of $0<\zeta<1$. The total spin Hall conductivity of the total Hall current ($J_{\mathrm{SH}}+J_{\mathrm{STH}}$) is larger than the conventional spin Hall conductivity. The spin Hall conductivity is enhanced by the STMC effect. The enhancement to the value of the spin Hall conductivity is most obvious at $\zeta\approx 1$. By comparing with the conventional value($e/16\pi$) of the spin Hall conductivity, the total spin Hall conductivity is enhanced by a factor of 2 at $\zeta\approx 1$.

It is difficult to distinguish the intrinsic spin Hall effect (ISHE) from the extrinsic spin Hall effect (ESHE) in the SOC system experimentally \cite{rev1}. However the numerical results show that the effect of STMC enhances the total spin Hall conductivity. The enhancement phenomenon of the spin Hall conductivity may provide an experimental method to distinguish the ISHE from the ESHE. For example, the 2D free fermi gas in the double sublattices structure is a good candidate to verify the effect of the enhancement in the experiments.

\section{Conclusion}

The new spin-tensor Hall current due to the coupling of the spin-tensor and the momentum is found in the spin continuity equation by using the Noether's theorem. We propose a model of 2DFS with the SOC and STMC. The spin Hall conductivity and the spin-tensor Hall conductivity of the ISHE are studied in the model. We find that the STMC effect enhances the total spin Hall transport of the ISHE. The enhancement is more evident with the increasing of the STMC strength. Our results may motivate further theoretical studies of ISHE.

\section*{Acknowledgements}

Y. P. Fu acknowledges the support of National Natural Science Foundation of China (No. 11805029) and the Yunnan Applied Fundamental Research Projects (No. 2017FD250). F. J. Huang acknowledges the support from Yunnan Local Colleges Applied Basic Research Projects (No. 2017FH001-112). Q. H. Chen acknowledges the support from National Natural Science Foundation of China (No. 11947404) and the Fundamental Research Funds for the Central Universities (No. 2682016CX078).

\appendix

\section{Relativistic Total-Angular Momentum Conservation and Spin Continuity Equation}

According to the Noether's theorem \cite{nother1}, the conservation of the total-angular momentum of a fermion described by a Lagrangian $\mathcal{L}=\mathcal{L}(x_{\mu},\theta_{\sigma},\partial_{\mu}\theta_{\sigma})$ is given by
\begin{eqnarray}
\partial_{\alpha}M^{\alpha\mu\nu}=0,
\label{conser1}
\end{eqnarray}
where $\partial_{\alpha}=\left(\frac{\partial}{c\partial t},\nabla\right)$. The total-angular momentum tensor contains the spin angular momentum tensor (SAMT) and orbit angular momentum tensor (OAMT), $M^{\alpha\mu\nu}=S^{\alpha\mu\nu}+L^{\alpha\mu\nu}$. The SAMT is defined as
\begin{eqnarray}
S^{\alpha\mu\nu}=  \frac{\partial\mathcal{L}}{\partial(\partial_{\alpha}\theta_{\sigma})}  I_{\sigma\rho}^{\mu\nu}\theta^{\rho},
\label{SAMT1}
\end{eqnarray}
where the coefficient $I_{\sigma\rho}^{\mu\nu}=$$\left(  1/4\right)
[\gamma^{\mu},\gamma^{\nu}]_{\sigma\rho}$ (for the fermion) and $g_{\sigma}^{\mu}g_{\rho}^{\nu}-g_{\rho}^{\mu}g_{\sigma}^{\nu}$ (for electromagnetic field) \cite{RQM,FQ}. The $\gamma$-matrices is represented as
\begin{eqnarray}
\gamma^{\mu}=\left(
                 \left(
                   \begin{array}{cc}
                     1 & 0 \\
                     0 & -1 \\
                   \end{array}
                 \right),
                 \left(
                   \begin{array}{cc}
                     0 & \boldsymbol{\sigma} \\
                     -\boldsymbol{\sigma} & 0 \\
                   \end{array}
                 \right)
\right),
\label{gama1}
\end{eqnarray}
$\boldsymbol{\sigma}$ is the Pauli matrices. $g^{\mu\nu}$=diag$(1,-1,-1,-1)$ is the Minkowski metric tensor $(\mu,\nu=0,1,2,3)$. The Greek letters denote the Lorentz indices, the Latin letters denote three-vector indices.

The OAMT can be written as
\begin{eqnarray}
L^{\alpha\mu\nu}=x^{\mu}T^{\alpha\nu}-x^{\nu}T^{\alpha\mu},
\label{OAMT1}
\end{eqnarray}
where $T^{\mu\nu}$ is the energy-momentum tensor
\begin{eqnarray}
T^{\mu\nu}=\frac{\partial\mathcal{L}}{\partial (\partial_{\mu}\theta_{\sigma})}\partial^{\nu}\theta_{\sigma}-g^{\mu\nu}\mathcal{L}.
\label{T1}
\end{eqnarray}

The field $\theta$ denotes the fermion or electromagnetic field ($\theta=(\Psi,A^{\mu})$). The system of Dirac fermions ($\Psi$) coupled with the electromagnetic filed ($A^{\mu}$) is described by the Lagrangian
\begin{eqnarray}
\mathcal{L}=\bar{\Psi}(i\hbar c\gamma^{\mu}
\partial_{\mu}-mc^{2})\Psi-e\bar{\Psi}\gamma^{\mu}A_{\mu}\Psi-\frac{1}{4}F^{\mu\nu}F_{\mu\nu},
\label{QED1}
\end{eqnarray}
where $F^{\mu\nu}=
\partial^{\mu}A^{\nu}-\partial^{\nu }A^{\mu}$ is the electromagnetic tensor, and $\bar{\Psi}=\Psi^{+}\gamma^{0}$. After the partial derivative calculations we have
\begin{eqnarray}
\partial_{\alpha}S^{\alpha\mu\nu} &=& \frac{i\hbar c}{4}\partial_{\alpha}\left(\bar{\Psi}\gamma^{\alpha}\left[\gamma^{\mu},\gamma^{\nu}\right]\Psi\right) +J^{\nu}A^{\mu}-J^{\mu}A^{\nu}\nonumber\\
&&+\left(\partial^{\alpha} A^{\nu}\right)\left(\partial_{\alpha} A^{\mu}\right)-\left(\partial^{\nu} A^{\alpha}\right)\left(\partial_{\alpha} A^{\mu}\right)\nonumber\\
&&-\left(\partial^{\alpha} A^{\mu}\right)\left(\partial_{\alpha} A^{\nu}\right)+\left(\partial^{\mu} A^{\alpha}\right)\left(\partial_{\alpha} A^{\nu}\right),
\label{SAMT2}
\end{eqnarray}
and
\begin{eqnarray}
\partial_{\alpha}L^{\alpha\mu\nu} &=& \bar{\Psi}\left(i\hbar c \gamma^{\mu}\partial^{\nu}-i\hbar c \gamma^{\nu}\partial^{\mu}\right)\Psi \nonumber\\
&&-\left(\partial^{\mu} A_{\alpha}\right)\left(\partial^{\nu} A^{\alpha}\right)+\left(\partial_{\alpha} A^{\mu}\right)\left(\partial^{\nu} A^{\alpha}\right)\nonumber\\
&&+\left(\partial^{\nu} A^{\alpha}\right)\left(\partial^{\mu} A_{\alpha}\right)-\left(\partial_{\alpha} A^{\nu}\right)\left(\partial^{\mu} A^{\alpha}\right),
\label{OAMT2}
\end{eqnarray}
where the electronic current is $J^{\mu}=\partial_{\nu} F^{\nu\mu}=e\bar{\Psi}\gamma^{\mu}\Psi$, and the energy-momentum tensor is $T^{\mu\nu}=\bar{\Psi}i\hbar c\gamma^{\mu}\partial^{\nu}\Psi-F^{\mu\rho} \partial^{\nu} A_{\rho}-g^{\mu\nu}\mathcal{L}$. The energy-momentum is conserved ($\partial_{\mu}T^{\mu\nu}=0$). By substituting (\ref{SAMT2}) and (\ref{OAMT2}) into the conservation equation of total-angular momentum, Eq. (\ref{conser1}) can be rewritten as
\begin{eqnarray}
\partial_{\alpha}\left(\frac{i\hbar c}{4}\bar{\Psi}\gamma^{\alpha}\left[\gamma^{\mu},\gamma^{\nu}\right]\Psi\right)=\bar{\Psi}c\left( \gamma^{\nu}\pi^{\mu}-\gamma^{\mu}\pi^{\nu}\right)\Psi,
\label{conser2}
\end{eqnarray}
here we define the momentum operator as $\pi^{\mu}=i\hbar \partial^{\mu}-\frac{e}{c}A^{\mu}$. If we only consider the vector indices of $\mu$ and $\nu$, one can easily derive
\begin{eqnarray}
&&\frac{\partial}{\partial t}\left(\Psi^{+}\varepsilon^{ijk}\frac{\hbar }{2}\Sigma_{k}\Psi\right)+\nabla_{l}\left(\Psi^{+}c\alpha^{l}\varepsilon^{ijm}\frac{\hbar }{2}\Sigma_{m}\Psi\right)=\nonumber\\
&&\Psi^{+}c\left(\pi^{i}\alpha^{j}-\pi^{j}\alpha^{i}\right)\Psi,
\label{conser3}
\end{eqnarray}
where $\varepsilon^{ijk}$ is the Levi-Civita symbol. Here the $\Sigma$ and $\alpha$ matrices are
\begin{eqnarray}
\boldsymbol{\Sigma}=\left(
                      \begin{array}{cc}
                        \boldsymbol{\sigma} & 0 \\
                        0 & \boldsymbol{\sigma} \\
                      \end{array}
                    \right),
\label{spin1}
\end{eqnarray}
and
\begin{eqnarray}
\boldsymbol{\alpha}=\left(
                      \begin{array}{cc}
                        0 & \boldsymbol{\sigma} \\
                        \boldsymbol{\sigma} & 0 \\
                      \end{array}
                    \right).
\label{spin1}
\end{eqnarray}

Since $\varepsilon^{ijm}\varepsilon_{ijn}=2\delta^{m}_{n}$ and $\varepsilon^{ijk} a_{i} b_{j}=(\boldsymbol{a}\times\boldsymbol{b})^{k}$, the relativistic continuity equation of the spin can be derived as
\begin{eqnarray}
\frac{\partial}{\partial t}\rho_{\mathrm{D}}^{i}
+\nabla_{j}J_{\mathrm{D}}^{ji}=T^{i}_{\mathrm{D}},
\label{conser2}
\end{eqnarray}
where $\rho_{\mathrm{D}}^{i}=\Psi^{+}s_{\mathrm{D}}^{i}\Psi$ is the spin density of the Dirac Fermion, $J_{\mathrm{D}}^{ji}=\Psi^{+}v^{j}_{\mathrm{D}}s_{\mathrm{D}}^{i}\Psi$ is the relativistic spin current originating from the contribution of the spin angular momentum, and $T^{i}_{\mathrm{D}}=\Psi^{+}\left(\boldsymbol{\pi}\times
\boldsymbol{v}_{\mathrm{D}}\right)^{i}\Psi$ is the spin torque originating from the contribution of the orbit angular momentum. Here we define the velocity operator as $\boldsymbol{v}_{\mathrm{D}}=c\boldsymbol{\alpha}$ and the spin operator as $\boldsymbol{s}_{\mathrm{D}}=\hbar\boldsymbol{\Sigma}/2$. Eq.(\ref{conser2}) is similar to the spin continuity equation of the SHE in Dirac-Rashba systems \cite{dirac-rashba}.








\end{document}